\documentclass[conference]{IEEEtran}
\IEEEoverridecommandlockouts
\usepackage{amsmath,amssymb,amsthm,mathtools}
\usepackage{bm}
\usepackage{graphicx}
\usepackage{booktabs}
\usepackage{cite}
\usepackage{url}
\usepackage{xcolor}
\usepackage{enumitem}
\usepackage{balance}
\usepackage{tikz}
\usetikzlibrary{arrows.meta,positioning,calc,decorations.pathreplacing,shapes.misc}
\usepackage[font=footnotesize]{caption}
\usepackage{algorithm}
\usepackage{algpseudocode}
\usepackage{xcolor}
\algrenewcommand\algorithmicrequire{\textbf{Input}}
\algrenewcommand\algorithmicensure{\textbf{Output}}
\usepackage[hidelinks]{hyperref}
\newcommand{\eqlabel}[1]{
	\text{ \setlength{\fboxsep}{0.1pt} \colorbox{gray!0}{
			\makebox[2cm][r]{\strut\textit{#1}} }}}
\usepackage[moderate,tracking=normal]{savetrees}
\usepackage{titlesec}
\titlespacing*{\section}{0pt}{0.1\baselineskip}{0.1\baselineskip}
\titlespacing*{\subsection}{0pt}{0.1\baselineskip}{0.1\baselineskip}
\newtheorem{theorem}{Theorem}
\newtheorem{proposition}{Proposition}
\newtheorem{corollary}{Corollary}
\newtheorem{remark}{Remark}

\newcommand{\m}{\boldsymbol}
\newcommand{\del}{\mathrm{del}}
\newcommand{\pk}{\mathrm{pk}}
\newcommand{\rs}{\mathrm{rs}}
\newcommand{\abs}{\mathrm{abs}}

\definecolor{fnavy}{HTML}{1D3558}
\definecolor{fpanel}{HTML}{EDF3F9}
\definecolor{fbox}{HTML}{D6E0E8}
\definecolor{forange}{HTML}{CD915B}
\definecolor{fgray}{HTML}{5E6E82}
\definecolor{fsand}{HTML}{F6E8DC}
\definecolor{fteal}{HTML}{196B69}
\definecolor{fcrimson}{HTML}{B83B3B}

\newif\ifsupp
\supptrue

\begin{document}

\title{\Large \textbf{\textit{The Race between Protection and Control:\\Fast Reactive Reserves Against Overvoltage Cascades}}}

\author{\IEEEauthorblockN{Abdallah Alalem Albustami, Mads Almassalkhi, and Ahmad F.~Taha}%
\thanks{A.~Alalem Albustami and A.~F.~Taha are with the Department of Civil and Environmental Engineering, Vanderbilt University, Nashville, TN, USA (emails \{abdallah.b.alalem.albustami, ahmad.taha\}@vanderbilt.edu). M.~Almassalkhi is with the Department of Electrical and Biomedical Engineering, University of Vermont, Burlington, VT, USA (email malmassa@uvm.edu).}%
\thanks{This work is supported by the National Science Foundation under Grant 2550934.}\vspace{-0.9cm}}

\maketitle

\begin{abstract}
In an overvoltage cascade, each relay trip removes reactive power absorption and raises the voltage at other plants, often faster than controls can act. This paper determines how much fast reactive power to hold, where, and how fast, so that no contingency trips a plant. A trip certificate allows a relay to pick up if it resets before its delay expires, making the sizing a linear program. A necessary condition identifies the relays that only preventive action can protect, and the reserve of one control decreases with the relay delay and increases with its latency. In nonlinear simulations of four systems, the corrective actions prevent every trip within the analysis window for relays without reset hysteresis, and each plant's own set point is the most effective reserve. A 2\% hysteresis or a small error in the relay voltage can change the reserve as much as the relay delay.\end{abstract}


\section{Introduction}\label{sec:intro}
\IEEEPARstart{I}{nverter}-based resources (IBRs) and large electronic loads such as data centers change their reactive power within milliseconds and are disconnected by their protection when the voltage leaves a band. About 1500~MW of data center load was lost in this way after a fault in July 2024~\cite{nerc_ir_2025}. Another incident, the Iberian blackout of 28 April 2025, motivates this work where fast changes of reactive power and protection resulted in blackouts affecting tens of millions of people. A voltage rise with plants operating at fixed power factor led to overvoltage trips of generation, each trip removed reactive power absorption and raised the voltage at other plants, and the system collapsed about 30~s after the first trip~\cite{final}. The final report concludes that manual voltage control was too slow in the last 38~s and that only dynamic voltage control could have limited the voltage~\cite[p.~328]{final}.

Control rooms have no tool that says how much of this fast control is needed. Contingency analysis checks steady-state flows and voltages~\cite{stott1987security}, voltage stability tools compute the distance to collapse~\cite{van2007voltage}, dynamic security assessment simulates a few contingencies~\cite{morison2004power}, and the reactive reserve is tracked as the unused output of dynamic devices~\cite{nerc2016rpp}. None of them uses the relay settings of the plants or compares the response time of the controls with the relay delays. The voltage that a plant relay measures, often on the low-voltage side of its transformer, is also not the one that the energy management system monitors~\cite[pp.~159--164]{final}. Ride-through requirements share this limitation, since the North American standard PRC-029 defines the region in which an IBR must stay connected, including 1~s above 1.10~pu, at the high side of the main power transformer~\cite{nerc2024prc029}, whereas a relay on the low-voltage side of a fixed-tap transformer can reach its pickup setting while the high-side voltage is still inside that region.

Reactive reserves have been sized for steady-state voltage stability, which is a low-voltage problem~\cite{capitanescu2011assessing}. For cascading trips of wind farms, voltage security regions keep the turbine voltages within their limits after one contingency~\cite{niu2016autonomous}, and the dynamic reactive reserve has been optimized to enlarge these regions~\cite{niu2018dynamic}. To our knowledge, these methods use the voltages after the contingency and not the relay delays or the latencies of the controls. A study of the Iberian event defines a static safety metric, the margin to an overvoltage-driven blackout~\cite{rouco2026overvoltage}. Our earlier work computes the voltage at every overvoltage relay within the relay time delay and finds fast corrective actions that keep every relay below pickup after one event, with the relay delay taken as given~\cite{albustami2026iberian}. It does not show how the reserve depends on the relay settings or what must be held against a contingency list.

This paper sizes the fast reactive reserve that prevents overvoltage cascades, by location and response time, and makes the following contributions. \textit{(i)} A trip certificate shows that a relay does not trip if every pickup is followed, within the relay delay, by an instant at which the voltage is below the reset level. The certificate is exact up to an error allowance and leads to a linear program (LP). \textit{(ii)} A necessary condition identifies the pairs of relay and event that no corrective action can protect and that need preventive action, and every pair is assigned to one of four classes. \textit{(iii)} For a single control, the reserve has a closed form that decreases with the relay delay and increases with the latency, and for all controls an LP over the contingency list gives the reserve at each location. \textit{(iv)} Case studies on four systems, one of which reproduces the mechanism of the Iberian blackout, quantify the reserve, show how much the reset hysteresis and an error in the relay voltage add to it, and validate the corrective actions in nonlinear simulation. With the method, an operator can check whether the reactive devices and plant set points in service protect every plant after, for example, the trip of a large plant, how much reserve to hold at each location, and which relay delays to lengthen.

\section{Model and Reserve Problem}\label{sec:model}
\subsection{Relay voltage model}\label{sec:responses}
\begin{figure}[t]
	\centering
	\resizebox{\columnwidth}{!}{%
		\begin{tikzpicture}[
			>=Stealth, font=\footnotesize,
			gblock/.style={rectangle,draw=none,fill=black!7,align=center,minimum height=9mm,inner sep=2pt},
			cblock/.style={rectangle,draw=none,fill=teal!20,align=center,minimum height=9mm,inner sep=2pt},
			pblock/.style={rectangle,draw=none,fill=red!25,align=center,minimum height=9mm,inner sep=2pt},
			sum/.style={circle,draw=none,fill=teal!100,minimum size=6mm,inner sep=0pt,font=\footnotesize,text=white},
			sig/.style={-Latex,black,thick},
			lab/.style={font=\small,inner sep=1pt},
			note/.style={font=\scriptsize,inner sep=1pt}
			]
			\node[gblock,minimum width=12mm] (E) at (0,0) {Event\\$j$};
			\node[sum,right=9mm of E] (S) {$\sum$};
			\node[gblock,minimum width=25mm,right=8mm of S] (N) {Network and\\plant controls};
			\node[pblock,minimum width=22mm,right=11mm of N] (R) {Overvoltage\\relay $i$};
			\draw[sig] (E) -- (S);
			\node[lab] at ([xshift=-2.6mm,yshift=2.6mm]S.west) {$+$};
			\draw[sig] (S) -- (N);
			\draw[sig] (N) -- node[lab,above] {$v_i(t)$} node[note,below,yshift=-0.6mm,text=red!60!black] {fixed tap} (R);
			\node[pblock,minimum width=25mm,below=2mm of N] (T) {Plant $i$ trips};
			\draw[sig] (R.south) |- (T.east);
			\draw[sig] (T.west) -| (S.south);
			\node[lab] at ([xshift=-2.6mm,yshift=-2.6mm]S.south) {$+$};
			\node[note,anchor=east,text=red!60!black] at ([xshift=-1.2mm]S.south |- T.west) {loss of absorption};
			\node[cblock,minimum width=12mm,above=4.75mm of E] (D) {Event\\detection};
			\node[cblock,minimum width=25mm] (C) at (N |- D) {Fast controls $k$};
			\draw[sig] (E) -- (D);
			\draw[sig] (D) -- node[lab,above] {$\m q^{(j)}$} (C);
			\draw[sig] (C.south) -- ++(0,-3.2mm) -| (S.north);
			\node[lab] at ([xshift=-2.6mm,yshift=2.6mm]S.north) {$-$};
			\node[note,text=teal!60!black,anchor=west] at ([xshift=1mm]C.east) {latency $\tau_k$};
	\end{tikzpicture}}
	\caption{Feedback loops of an overvoltage cascade. A relay trip removes the reactive power absorption of its plant and raises the voltage at the other relays (red). Fast controls, armed by the detection of the event, absorb the reactive power $\m q^{(j)}$ after their latencies and lower the relay voltages (teal).}
	\label{fig:feedback}
\end{figure}
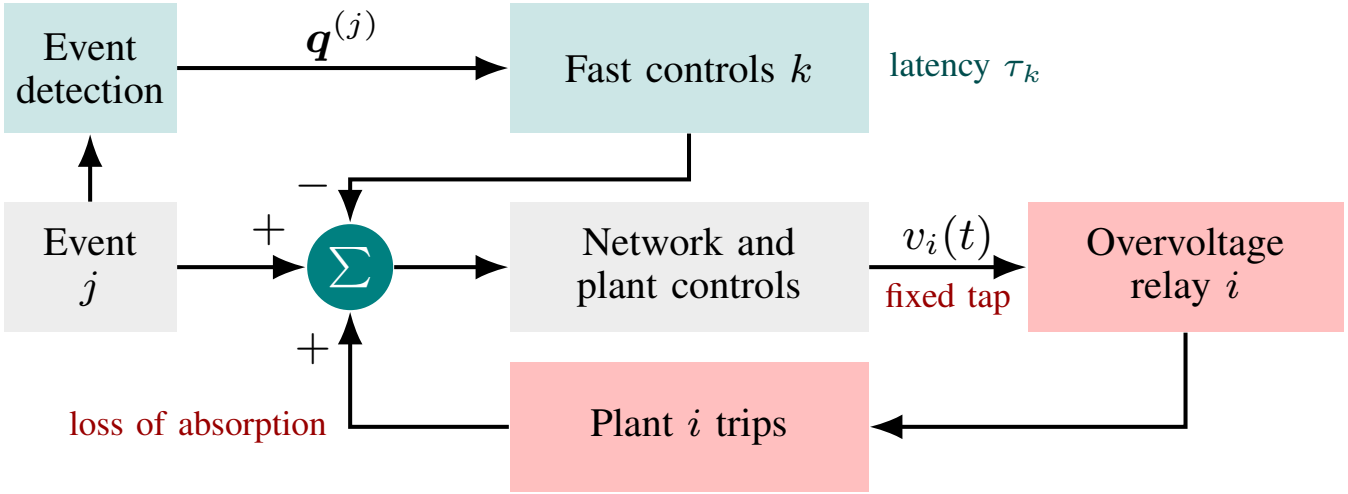

Consider plants $i\in\mathcal I$, each protected by an overvoltage relay, events (contingencies) $j\in\mathcal J$, and fast controls $k\in\mathcal K$, which interact as shown in Fig.~\ref{fig:feedback}. The relay of plant $i$ measures the voltage $v_i$ of the collector bus or of the low-voltage side of the main power transformer, with a pickup setting $V_i^{\pk}$, a definite-time delay $T_i^{\del}$, and a reset (dropout) ratio $r_i\le1$. Let $v_i^0$ be the relay voltage before the event. The margin to pickup is $h_i=V_i^{\pk}-v_i^0>0$, and the margin to the reset level is $h_i^{\rs}=r_iV_i^{\pk}-v_i^0=h_i-(1-r_i)V_i^{\pk}$, which is negative when the hysteresis band $(1-r_i)V_i^{\pk}$ exceeds $h_i$.

Within a window $[0,T]$ of a few seconds after an event, with $T\ge T_i^{\del}$, taps, limiters, and control modes do not change, and the relay voltages follow the differential-algebraic model of the grid linearized at the operating point. A step or ramp event enters this model as a disturbance. A switching event changes the network, and the jump of the algebraic variables is computed exactly with the dynamic states held at their values before the switch~\cite{albustami2026iberian}. Let $p_{ij}(t)$ be the rise of $v_i$ after event $j$ without control, in per unit. Control $k$ absorbs the reactive power $q_k\in[0,q_k^{\max}]$, in MVAr, through a response that is zero before its latency $\tau_k>0$, which includes detection, communication, and actuation. Let $g_{ik}(t)$ be the resulting reduction of $v_i$ per MVAr of control $k$, in pu/MVAr, computed for the post-contingency network (the index $j$ is omitted). By superposition, the relay voltage after event $j$ under the corrective action $\m q=(q_k)_{k\in\mathcal K}$ is
\begin{equation}\label{eq:closed}
v_{ij}(t;\m q)=v_i^0+p_{ij}(t)-\textstyle\sum_{k\in\mathcal K}g_{ik}(t)\,q_k .
\end{equation}
Throughout, $t=0$ is the instant just after the event, so $p_{ij}(0)$ is the jump that the network alone sets and that no control can change. We refer to relay $i$ after event $j$ as the pair $(i,j)$.

\subsection{Relay logic}\label{sec:relay}
Relay $i$ picks up when $v_i\ge V_i^{\pk}$. Its timer then runs while $v_i\ge r_iV_i^{\pk}$ and resets to zero when $v_i$ falls below $r_iV_i^{\pk}$, and the relay trips when the timer reaches $T_i^{\del}$. A reset ratio of 0.98 at a pickup of 1.10~pu gives a band of 0.022~pu, comparable to the margins of plants in a high-voltage state. When the band exceeds the margin, $h_i^{\rs}<0$, and the voltage must fall below its value before the event for the timer to reset. No relay is picked up before the event, but an event may pick a relay up at $t=0$. The relay voltage is known only up to the error of the linear model and of the estimate of $v_i^0$. Let $\delta_i>0$ bound this error. The certificates below lower both the pickup and the reset level by this allowance, so a relay with $\delta_i>h_i$ cannot be certified even without an event. 

\subsection{Reserve problem}\label{sec:problem}
\begin{figure}[t]
	\centering
	\resizebox{\columnwidth}{!}{%
		\begin{tikzpicture}[
			x=1mm, y=1mm, >=Stealth, font=\footnotesize,
			nohyph/.style={execute at begin node={\hyphenpenalty=10000\exhyphenpenalty=10000}},
			box/.style={nohyph,rectangle,draw=none,align=center,inner xsep=1mm,inner ysep=1.5pt},
			gblock/.style={box,fill=black!7},
			cblock/.style={box,fill=teal!20},
			pblock/.style={box,fill=red!25},
			w4/.style={text width=18.5mm,minimum height=10mm},
			w4s/.style={text width=18.5mm,minimum height=13mm},
			w2/.style={text width=41mm,minimum height=10mm},
			num/.style={circle,draw=none,fill=teal!100,text=white,font=\scriptsize\bfseries,minimum size=4mm,inner sep=0pt},
			sig/.style={-Latex,black,thick},
			ref/.style={font=\scriptsize}
			]
			\node[gblock,w4] (i1) at (-33.75,0) {Operating\\point};
			\node[gblock,w4] (i2) at (-11.25,0) {Relay settings\\$V_i^{\pk}$, $T_i^{\del}$, $r_i$};
			\node[gblock,w4] (i3) at (11.25,0) {Contingency\\list};
			\node[gblock,w4] (i4) at (33.75,0) {Fast controls\\$q_k^{\max}$, $\tau_k$};
			\foreach \n in {i1,i2,i3,i4} \draw[thick] (\n.south) -- ++(0,-1.5);
			\draw[thick] (-33.75,-6.5) -- (33.75,-6.5);
			\draw[sig] (-33.75,-6.5) -- (-33.75,-9.5);
			\node[cblock,w4s,anchor=north] (s1) at (-33.75,-9.5) {\textbf{Voltage}\\\textbf{responses}\\{\scriptsize Section~\ref{sec:responses}}};
			\node[cblock,w4s,anchor=north] (s2) at (-11.25,-9.5) {\textbf{Trip}\\\textbf{certificate}\\{\scriptsize Theorem~\ref{thm:cert}}};
			\node[pblock,w4s,anchor=north] (s3) at (11.25,-9.5) {\textbf{Necessary}\\\textbf{condition}\\{\scriptsize Proposition~\ref{prop:necessary}}};
			\node[cblock,w4s,anchor=north] (s4) at (33.75,-9.5) {\textbf{Reserve}\\\textbf{LP}\\{\scriptsize Section~\ref{sec:reserve}}};
			\draw[sig] (s1) -- (s2);
			\draw[sig] (s2) -- (s3);
			\draw[sig] (s3) -- (s4);
			\node[num] at (s1.north west) {1};
			\node[num] at (s2.north west) {2};
			\node[num,fill=red!70!black] at (s3.north west) {3};
			\node[num] at (s4.north west) {4};
			\node[gblock,w2,anchor=north,minimum height=11mm] (o1) at (-22.5,-27) {\textbf{Pairs that need preventive action}\\{\scriptsize longer delay, tap change,}\\[-1pt]{\scriptsize or no switching}};
			\node[gblock,w2,anchor=north,minimum height=11mm] (o2) at (22.5,-27) {\textbf{Reserve $\bar{\m q}$ and actions $\m q^{(j)}$}\\{\scriptsize MVAr held at each control}\\[-1pt]{\scriptsize and used after each event}};
			\draw[sig] (s3.south) -- ++(0,-2) -| (o1.north);
			\draw[sig] (s4.south) -- (s4.south |- o2.north);
	\end{tikzpicture}}
	\caption{Inputs, steps, and outputs of the method.}
	\label{fig:flow}
\end{figure}
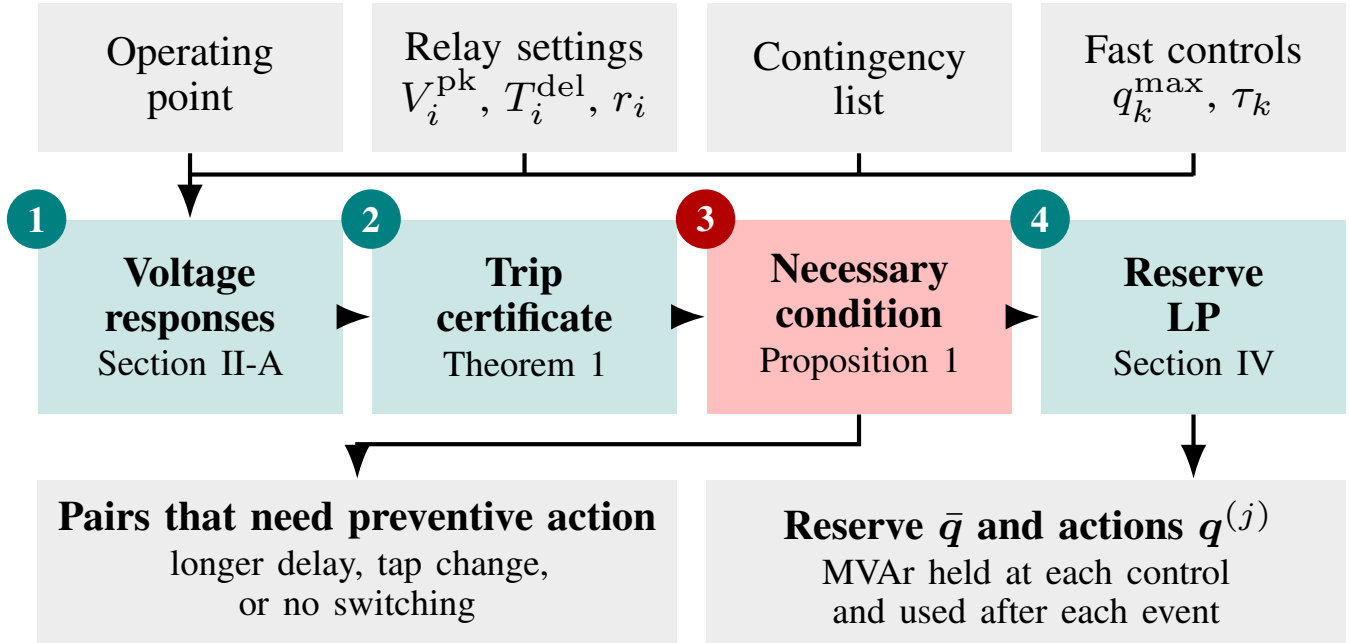

Given the contingency list, the relay settings, and the controls, the problem is to find the smallest fast reactive reserve $\bar{\m q}$, the MVAr held at each control, such that for every event $j$ some corrective action $\m q^{(j)}\le\bar{\m q}$ is certified to prevent every trip within the window. The parameters of interest are the relay delays $T_i^{\del}$, which the operator can ask of the plant owners through the coordination of protection settings~\cite{eu2016rfg,nerc2024prc029}, and the latencies $\tau_k$, which depend on the type, location, and triggering of each control. Fig.~\ref{fig:flow} summarizes the method.

\section{Trip Certificate and Bounds on the Reserve}\label{sec:race}
\subsection{Trip certificate}\label{sec:certificate}
The pickup certificate of~\cite{albustami2026iberian} keeps the relay voltage below pickup at every instant, which is impossible when the jump $p_{ij}(0)$ exceeds the margin $h_i$. The relay logic asks for less, since a pickup does no harm if the relay resets before its delay ends.

\begin{theorem}[Trip certificate]\label{thm:cert}
	Let $t_1<\cdots<t_N$ be reset instants in $(0,T]$, and let $t_0=0$ and $t_{N+1}=T$. If
	\begin{subequations}\label{eq:cert}
		\begin{align}
			v_{ij}(t_n;\m q)&\le r_iV_i^{\pk}-\delta_i,\quad n=1,\ldots,N,\label{eq:cert_reset}\\
			v_{ij}(t;\m q)&\le V_i^{\pk}-\delta_i,\quad t\in[t_n,\,t_{n+1}-T_i^{\del}],\ n=0,\ldots,N,\label{eq:cert_pickup}
		\end{align}
	\end{subequations}
	then relay $i$ does not trip in $[0,T]$ after event $j$ under the action $\m q$. Conversely, every action under which the relay does not trip when both levels are lowered by more than $\delta_i$ satisfies~\eqref{eq:cert} for some reset instants.
\end{theorem}
The proof is given in the supplementary material (SM).\footnote{The SM is included in the extended version of this paper, which is available online under the same title, also available in the repo~\cite{albustami2026code}} Condition~\eqref{eq:cert_reset} asks the voltage to be below the reset level at each reset instant, and~\eqref{eq:cert_pickup} asks it to stay below pickup wherever no reset instant follows within the delay, so every pickup is followed by a reset before the timer expires. With no reset instant,~\eqref{eq:cert} is the pickup certificate on $[0,T-T_i^{\del}]$.

\subsection{Necessary condition}\label{sec:necessary}
If the jump exceeds the margin, $p_{ij}(0)\ge h_i$, the relay picks up at $t=0$, before any control acts, and the controls have exactly $T_i^{\del}$ to reset it. Let $\underline p_{ij}=\inf_{0<t\le T_i^{\del}}p_{ij}(t)$ be the smallest rise without control within the delay and $\bar g_{ik}=\sup_{0<t\le T_i^{\del}}[g_{ik}(t)]_+$ the largest effect of control $k$ within it, where $[x]_+=\max\{x,0\}$.

\begin{proposition}[Necessary condition]\label{prop:necessary}
Suppose that $p_{ij}(0)\ge h_i$. Every action $\m q\ge\m0$ that prevents the trip of relay $i$ in $[0,T]$ satisfies
\begin{equation}\label{eq:necessary}
\textstyle\sum_{k\in\mathcal K}\bar g_{ik}\,q_k>\underline p_{ij}-h_i^{\rs}.
\end{equation}
If~\eqref{eq:necessary} fails for $\m q=\m q^{\max}$, no admissible action prevents the trip.
\end{proposition}
The proof is given in the SM.

Each pair $(i,j)$ at risk, that is, whose relay picks up without control, belongs to one of four classes. The pair is secure if~\eqref{eq:cert} holds with $\m q=\m0$. It requires preventive action if no corrective action of any size satisfies~\eqref{eq:cert}, because no control lowers the voltage at the instants where~\eqref{eq:cert} is violated. The pairs that fail~\eqref{eq:necessary} for every capability are of this kind, and only a measure taken before the event helps them, such as a tap change, a longer relay delay, or avoiding the switching. The pair is correctable if some corrective action $\m q\le\m q^{\max}$ satisfies~\eqref{eq:cert}, which a mixed-integer program over the reset instants decides exactly for one pair, and not correctable with the available controls otherwise. Since the certificate is exact up to $\delta_i$, the last class means that the available controls cannot prevent the trip in the linear model with that allowance.

\subsection{Reserve of a single control}\label{sec:one}
For one control, or a fixed mix of controls applied in proportion, the smallest action that prevents the trip has a closed form. Let $g_{ik}\ge0$ on $(0,T]$ and define the reactive power that holds relay $i$ at pickup and at the reset level at time $t$,
\begin{equation}\label{eq:a}
a_{ij}^{\pk}(t)=\frac{p_{ij}(t)-h_i}{g_{ik}(t)},\qquad a_{ij}^{\rs}(t)=\frac{p_{ij}(t)-h_i^{\rs}}{g_{ik}(t)},
\end{equation}
with the convention $x/0=+\infty$ for $x\ge0$ and $x/0=-\infty$ for $x<0$.

\begin{proposition}[Reserve of one control]\label{prop:one}
Control $k$ alone prevents the trip of relay $i$ in $[0,T]$ after event $j$ if and only if $q_k>q_{ij}^\star$, where
\begin{equation}\label{eq:qstar}
q_{ij}^\star=\max_{0\le s\le T-T_i^{\del}}\ \min\Big\{a_{ij}^{\pk}(s),\ \min_{s\le t\le s+T_i^{\del}}a_{ij}^{\rs}(t)\Big\}.
\end{equation}
The reserve $[q_{ij}^\star]_+$ is nonincreasing in $T_i^{\del}$. If the unit response of the control is a nondecreasing function of $t-\tau_k$, the reserve is nondecreasing in $\tau_k$.
\end{proposition}
The proof is given in the SM.

Two shapes of the rise without control are common. When the voltage jumps and then stays nearly constant, the first window of~\eqref{eq:qstar} binds, and once $T_i^{\del}$ exceeds the time the control needs to reach its full effect, a longer delay lowers the reserve little, because the same action must still reset the relay. When the voltage returns below the reset level on its own within a time $L$, the reserve falls as $T_i^{\del}$ approaches $L$ and vanishes beyond it, so the delay, not the latency, sets the reserve. For a jump, the reserve grows with the lost reactive power absorption and is inversely proportional to the effect of the control at the relay (SM, Corollary~S1). This effect is largest for the plant's own reactive power set point, because its reactive power also flows through the plant transformer.

\section{Sizing the Fast Reactive Reserve}\label{sec:reserve}
\subsection{Reserve linear program}\label{sec:lp}
The reserve must cover every relay and every event of the list at once, with the controls that remain available after each event. Let $\mathcal S_j$ be the plants that event $j$ disconnects, $\mathcal K_j\subseteq\mathcal K$ the controls available after it, which excludes the set points of the plants in $\mathcal S_j$ and the set points of tripped units, and $\mathcal R_j$ the relays at risk after event $j$, except those that require preventive action. For $i\in\mathcal R_j$, let $\mathcal T_{ij}$ be the reset instants and $\mathcal P_{ij}$ the time steps at which~\eqref{eq:cert_pickup} is imposed. The variables are the reserve $\bar{\m q}$ held at each control, the corrective action $\m q^{(j)}$ after each event, and a slack $\eta_{ij}\ge0$ for each pair, in units of the margin $h_i$. For every event $j\in\mathcal J$, the constraints are
\begin{subequations}\label{eq:lp}
	\begin{align}
		& v_{ij}(t;\m q^{(j)})\le r_iV_i^{\pk}-\delta_i+h_i\eta_{ij},\ \ t\in\mathcal T_{ij},\ i\in\mathcal R_j,\label{eq:lp_reset}\\
		& v_{ij}(t;\m q^{(j)})\le V_i^{\pk}-\delta_i+h_i\eta_{ij},\ \ t\in\mathcal P_{ij},\ i\in\mathcal R_j,\label{eq:lp_pickup}\\
		& v_{ij}(t;\m q^{(j)})\le\max\{v_i^0+p_{ij}(t),\,V_i^{\pk}-\delta_i\},\ \ i\notin\mathcal R_j\cup\mathcal S_j,\label{eq:lp_noworse}\\
		& \m0\le\m q^{(j)}\le\bar{\m q}\le\m q^{\max},\ \ q^{(j)}_k=0\ \text{for}\ k\notin\mathcal K_j,\ \ \eta_{ij}\ge0.\label{eq:lp_bounds}
	\end{align}
\end{subequations}
Constraints~\eqref{eq:lp_reset} and~\eqref{eq:lp_pickup} are the certificate~\eqref{eq:cert} relaxed by the slack, and constraints~\eqref{eq:lp_pickup} use bounds on the responses that hold between time steps. Constraints~\eqref{eq:lp_noworse} hold at every time step and keep an action from raising a relay that is not at risk above both its voltage without control and its lowered pickup level. A pair with $\eta_{ij}=0$ is covered, that is, by Theorem~\ref{thm:cert} its relay does not trip under $\m q^{(j)}$. 
The LP is solved in three stages,
\begin{subequations}\label{eq:stages}
	\begin{alignat}{2}
		&\hspace{-0.9cm}\eqlabel{\textbf{Coverage}}\;\;\;\;
		&&\eta^\star=\min\ \textstyle\sum_{i,j}\eta_{ij}\ \ \text{s.t.~\eqref{eq:lp}},
		\label{eq:stage1}\\
		&\hspace{-0.9cm}\eqlabel{\textbf{Reserve}}
		&&\bar{\m q}^\star=\arg\min\ \textstyle\sum_{k}\bar q_k\ \ \text{s.t.~\eqref{eq:lp}},\ \textstyle\sum_{i,j}\eta_{ij}\le\eta^\star,
		\label{eq:stage2}\\
		&\hspace{-0.9cm}\eqlabel{\textbf{Actions}}
		&&\m q^{(j)\star}=\arg\min\ \textstyle\sum_{k}q_k^{(j)}\ \ \text{s.t.~\eqref{eq:lp}},\ \bar{\m q}=\bar{\m q}^\star,
		\label{eq:stage3}
	\end{alignat}
\end{subequations}
where~\eqref{eq:stage3} is solved for each event $j$ separately and keeps covered the pairs that~\eqref{eq:stage2} covers. Any reset instants give a valid certificate, so the reserve $\sum_k\bar q_k^\star$ is an upper bound on the smallest reserve. The corrective actions depend on the event, which presumes an event-based remedial action scheme (RAS), armed for example by breaker status. A scheme that does not identify the event can apply one action to every event, or, as a response-based scheme, let each plant act when its own relay picks up, and Section~\ref{sec:cases_scheme} compares these options. A relay with several definite-time stages trips when any of its stages trips, so Theorem~\ref{thm:cert} applies to each stage with its own pickup, delay, and reset ratio, and each stage has its own constraints~\eqref{eq:lp_reset} and~\eqref{eq:lp_pickup} with its own reset instants. Solving the LP for a range of relay delays gives the reserve for given settings and, read the other way, the shortest delay for which a given reserve suffices, which is what the coordination of protection settings needs.

\subsection{Choice of the reset instants}\label{sec:dp}
Constraints~\eqref{eq:lp_reset} and~\eqref{eq:lp_pickup} depend on the reset instants, which are chosen for each pair along the current action $\m d$ of its event. A dynamic program over the time steps finds the reset instants that minimize the largest multiple of $\m d$ that~\eqref{eq:cert} requires, and for one control it returns~\eqref{eq:qstar}. This choice alternates with stages~\eqref{eq:stage1} and~\eqref{eq:stage2} until the result stops improving, judged first by the uncovered pairs, then by the total slack, and then by the reserve. The iteration starts from the action of the pickup certificate and from reset instants spaced closer than the delay, keeps the better result, and ends with~\eqref{eq:stage3} and a check of every covered pair against~\eqref{eq:cert} (SM, Algorithm~S1).

\section{Case Studies}\label{sec:cases}
The method is applied to three benchmarks of~\cite{albustami2026iberian}, the IEEE 39-bus system, NPCC, and a 2224-bus model of Great Britain (GB), and to the stressed state of the 2000-bus system of~\cite{albustami2026iberian}, built from~\cite{birchfield2016grid}, which reproduces the mechanism of the Iberian overvoltage cascade. The margins of these states are small by design, from 0.008 to 0.08~pu, so that many relays are at risk (SM, Table~S1). Every plant has an overvoltage relay at its collector bus with a pickup of 1.10~pu and a delay between 0 and 1~s, which resets at pickup unless stated otherwise. The 1-s delay is the ride-through time of~\cite{nerc2024prc029} above 1.10~pu, and delays of 0.16~s and less represent faster equipment protection. As in~\cite{albustami2026iberian}, static synchronous compensators (STATCOMs) act after 20~ms with a lag of 30~ms, plant reactive power set points after 100~ms, AVR set points after 50~ms, thyristor-controlled reactors (TCRs) after 150~ms, and manually switched shunt reactors after 30~s. The headroom of each plant set point is the additional absorption allowed by a power factor of 0.9 at the plant rating, and the TCRs are modeled as continuously controlled susceptances. AVR set points, whose command is a reduction of the voltage set point, enter the classes but not the reserve, which counts MVAr. In the 2000-bus system, each plant has its own set point and a 100-MVAr STATCOM and TCR at its transmission bus, and the 12 units of the area have AVR set points. The responses are computed over $T=2$~s with a time step of 5~ms and $\delta_i=0.05h_i$, and the actions are simulated in ANDES~\cite{andes} with the relays. The SM reports the effect of the time step, the runtime, and a comparison with exact mixed-integer programs, and the code is publicly available~\cite{albustami2026code}.

\begin{figure}[t]
\centering
\includegraphics[width=0.93\columnwidth]{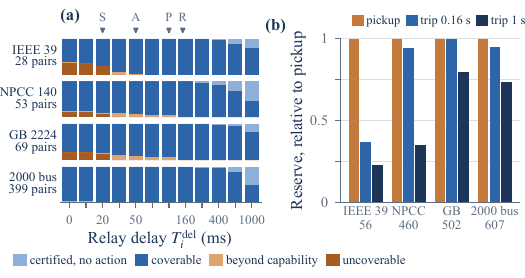}
\caption{Relay delay and reserve in the four systems. (a) Pairs at risk by class versus the relay delay, with the latencies of STATCOMs (S), AVR set points (A), plant set points (P), and TCRs (R). (b) Reserve of the trip certificate with 0.16-s and 1-s relays relative to that of the pickup certificate~\cite{albustami2026iberian} on the pairs that both can cover, with the reserve of the pickup certificate in MVAr.}
\label{fig:delay}
\end{figure}

\subsection{Effect of the relay delay}\label{sec:cases_delay}
Relays faster than the fastest control leave pairs that no reserve can protect, whereas from 0.16~s on the available controls cover every pair. Fig.~\ref{fig:delay}(a) gives the classes of Section~\ref{sec:necessary} for the pairs at risk, 28 to 399 per system, and delays from 0 to 1~s. At delays up to the 20-ms latency of the fastest control, 6 to 15 pairs per system require preventive action, as Proposition~\ref{prop:necessary} predicts. Between 35 and 100~ms, up to 10 pairs per system are not correctable with the available controls. From 160~ms on, every pair is correctable, and at 1~s, 255 of the 549 pairs are secure because the voltage returns below the reset level within the delay. On the IEEE 39-bus system, the reserve at 50~ms takes all 300~MVAr of the STATCOMs and still leaves one pair uncovered, whereas 48~MVAr of plant set points cover every pair at 0.16~s. Fig.~\ref{fig:delay}(b) compares the two certificates on the pairs that both can cover. The trip certificate needs up to 63\% less reserve with 0.16-s relays and 20 to 77\% less with 1-s relays, and it also covers 10 to 15 pairs per system that the pickup certificate cannot cover.

\begin{figure}[t]
\centering
\includegraphics[width=0.93\columnwidth]{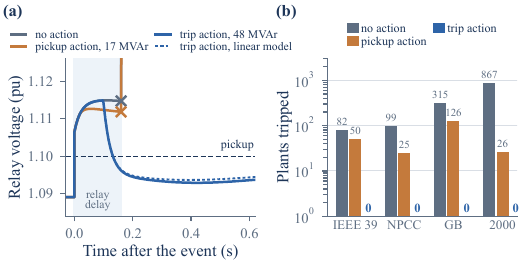}
\caption{Validation in ANDES. (a) Relay voltage of plant 7 of the IEEE 39-bus system after a load disconnection with 0.16-s relays, with crosses at the trips. (b) Plants tripped within 2~s over all events with a pair at risk, with 0.16-s and 1-s relays, counted until the simulation stops in the 24 runs that collapse.}
\label{fig:validation}
\end{figure}

\subsection{Validation in nonlinear simulation}\label{sec:cases_validation}
With relays that reset at pickup, the corrective actions of the trip certificate prevent every trip within the window in nonlinear simulation. Fig.~\ref{fig:validation}(a) shows the IEEE 39-bus system after a load disconnection that lifts the relay of plant 7 above its pickup at once. With 0.16-s relays and no action, the relay trips at the end of its delay and nine more plants follow. The pickup certificate cannot cover this relay, so its action of 17~MVAr leaves the same ten trips. The action of the trip certificate, 48~MVAr of plant set points, lets the relay pick up and resets it at 0.13~s, 28~ms before its timer expires, as the linear model predicts, and no plant trips. Over all 108 events with a pair at risk, Fig.~\ref{fig:validation}(b), the actions of the trip certificate prevent every trip within the window with 0.16-s and 1-s relays, whereas 1363 trips occur without action. The actions of the pickup certificate leave 227 trips, because the plants that it cannot cover trip and their lost reactive power absorption trips others. With 50-ms relays, the events with a pair that is not correctable still trip, as predicted, and no covered relay trips before an uncovered one.

\subsection{Effect of the reset ratio}\label{sec:cases_reset}
A reset hysteresis raises the reserve, and actions sized for the wrong reset ratio do not prevent the trips. With a reset ratio of 0.98, the hysteresis band of 0.022~pu is 0.28 to 2.75 times the margin across the plants, and the reserve with 0.16-s relays grows by a factor of 1.03 to 3.9 (SM, Table~S1). On IEEE 39 and GB, this increase exceeds what lengthening the delay from 0.16 to 1~s saves, and on NPCC and the 2000-bus system it is smaller. In ANDES, relays with this reset trip under the actions sized for a reset at pickup, 53 times on IEEE 39 and 33 times on NPCC with 0.16-s relays. The actions sized for the 0.98 reset prevent every trip on IEEE 39 at both delays and on NPCC with 0.16-s relays. On NPCC with 1-s relays, one covered relay trips, because its certificate rests on one late reset instant at which the simulated voltage exceeds the linear model by 0.14 times the margin, more than the allowance. GB and the 2000-bus system were not simulated with this reset. The reset ratio of each relay must therefore be known as well as its pickup and its delay.

\begin{figure}[t]
\centering
\includegraphics[width=0.93\columnwidth]{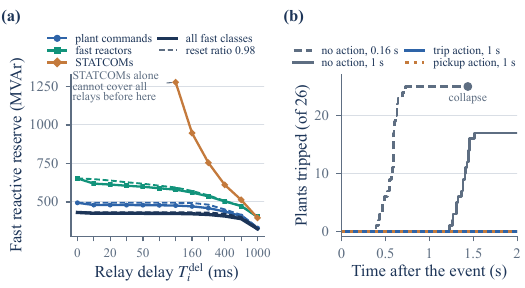}
\caption{Documented output reduction in the 2000-bus system. (a) Reserve that keeps all 26 plants connected, by type of control, with reset at pickup (solid) and at 0.98 of pickup (dashed). (b) Plants tripped in ANDES.}
\label{fig:activsg}
\end{figure}

\subsection{Type, latency, and location of the controls}\label{sec:cases_insights}
The plants' own reactive power set points are the most effective reserve, and the latency matters most when an event lifts the voltage at once. Fig.~\ref{fig:activsg}(a) gives the reserve that keeps all 26 plants of the 2000-bus system connected after the documented output reduction of 708~MW at fixed power factor~\cite{albustami2026iberian}. Plant set points need 494~MVAr with instantaneous relays and 330~MVAr with 1-s relays, and TCRs at the transmission buses need 15 to 32\% more. The STATCOMs, 100~MVAr each, cannot cover all relays below 100~ms, and the AVR set points of the 12 units never cover all relays. In ANDES, Fig.~\ref{fig:activsg}(b), 17 plants trip within 2~s with 1-s relays and no action, as in~\cite{albustami2026iberian}, and 25 trip before the system collapses with 0.16-s relays. The action for this event within the reserve of all events, 331~MVAr with 1-s relays, and the action of the pickup certificate, 697~MVAr, both keep every plant connected.

The latency matters most after a jump (SM, Fig.~S1). For the load disconnection of Fig.~\ref{fig:validation}(a), the smallest plant set point for this event alone stays at 31~MVAr with 0.16-s relays until the latency reaches 120~ms, triples at 140~ms, and no reserve covers the event once the latency reaches the delay, as Proposition~\ref{prop:necessary} states. The 48~MVAr of Fig.~\ref{fig:validation}(a) is larger because the reserve of all events holds less at the most effective plant. For the slow rise after the trip of the Seville plant of the 2000-bus system, the reserve grows by only 5\% between latencies of 0 and 300~ms, whereas 1-s relays instead of 0.16-s relays cut it by 62\%. A plant's own set point lowers its relay voltage by 0.01~pu with a median of 4.2 times less MVAr than a STATCOM at its transmission bus.

\subsection{Sensitivity to the relay voltage and to the window}\label{sec:cases_limits}
The reserve is needed mainly in stressed states, and it is only as reliable as the relay voltages and the linear model. With margins of at least 0.04~pu, the 2000-bus system has no pair at risk, and with margins of at least 0.05~pu, the three benchmarks have 0, 5, and 6. An error of 0.005~pu in the relay voltage, which is 0.06 to 0.62 times the margin here, raises the reserve with 0.16-s relays by a factor of 1.04 to 3.1 (SM, Table~S1). Actions sized without this error leave 81 trips on IEEE 39 and 402 on the 2000-bus system when the pickup is 0.005~pu lower than assumed. The allowance $\delta_i$ covers such errors at a cost, which is a factor of 1.8 on IEEE 39 for an error of 0.005~pu.

The certificate holds only within its window. Simulated for 30~s with the action held, 16 of the 21 actions tested leave no trip, and the other five let plants trip after the window. A window of 15~s with its last reset instant at $T$ removes these trips where the voltages settle, as on IEEE 39 (SM, Fig.~S2(a)), but it cannot certify the pairs whose voltage does not settle, 7 of 29 on IEEE 39. On NPCC, whose stressed state is unstable after most events, the simulated voltage exceeds the linear model by up to 0.024~pu (SM, Fig.~S2(b)), more than the margin of 6 of its 16 relays, and no event can be certified over the longer window.

\subsection{Schemes that do not identify the event}\label{sec:cases_scheme}
One action applied to every event, which removes the need to identify the event, needs at most 2\% more reserve on IEEE 39, NPCC, and the 2000-bus system and 3.2 times as much on GB, but applied in full after events that need little, it lowers bus voltages to 0.39~pu on the 2000-bus system, where the smallest actions of each event keep every bus above 0.94~pu. With a response-based rule in which each plant applies its full absorption 100~ms after its own relay picks up, no plant trips in 30~s on IEEE 39 and NPCC at either delay, including in the events that the certificate cannot cover, with 1.6 to 1.8 times the reactive power of the smallest actions on IEEE 39 and one NPCC bus at 0.89~pu. Other response-based rules, and the same rule limited to the reserve that the LP holds at each plant, leave trips (SM, Table~S2). The reserve should therefore be sized with an allowance for the voltage error and over a window in which the voltages settle.

\section{Conclusion}\label{sec:conclusion}
This paper sized the fast reactive reserve against overvoltage cascades, with the relay delay as the time the controls have to reset a relay that an event has picked up. A linear program gives the reserve at each location, and a necessary condition identifies the pairs that need preventive action. In four systems, relays faster than the fastest control left such pairs, whereas 0.16-s relays let the available controls cover every pair, and with relays that reset at pickup the corrective actions prevented every trip within the analysis window in nonlinear simulation. The plants' own reactive power set points were the most effective reserve. The reset hysteresis and a small error in the relay voltage can change the reserve as much as the relay delay, so relay settings and voltages must be verified before a reserve is counted. The certificate holds for the linear model within its allowance and its window, and where it does not hold, plants that act on their own relay voltage can back it up. The phasor models used here omit electromagnetic transients, inverter current loops, and relay measurement windows, which can affect the results below about 100~ms. Inverse-time relays, the handover to slower controls, and verification in electromagnetic transient simulation are left for future work.

\section*{Acknowledgment}
Anthropic Claude, an AI tool, wrote the implementation code and assisted with the proofs and the editing of the text. The authors reviewed the code and the derivations and take full responsibility for the content.

\balance
\bibliographystyle{IEEEtran}
\bibliography{refs}

\ifsupp
\setcounter{algorithm}{0}
\renewcommand{\thealgorithm}{S\arabic{algorithm}}
\setcounter{corollary}{0}
\renewcommand{\thecorollary}{S\arabic{corollary}}
\setcounter{section}{0}
\renewcommand{\thesection}{S\arabic{section}}
\setcounter{equation}{0}
\renewcommand{\theequation}{S\arabic{equation}}
\renewcommand{\theHsection}{S\arabic{section}}
\renewcommand{\theHequation}{S\arabic{equation}}
\setcounter{figure}{0}
\renewcommand{\thefigure}{S\arabic{figure}}
\renewcommand{\theHfigure}{S\arabic{figure}}
\setcounter{table}{0}
\renewcommand{\thetable}{S\arabic{table}}
\renewcommand{\theHtable}{S\arabic{table}}

\twocolumn[{\begin{center}
	{\large\bfseries Supplementary Material}\\[3pt]
	{\bfseries The Race between Protection and Control: Fast Reactive Reserves Against Overvoltage Cascades}\\[2pt]
	Abdallah Alalem Albustami, Mads Almassalkhi, and Ahmad F.~Taha
\end{center}\vspace{4pt}}]
This supplement gives the proofs of the results of Section~\ref{sec:race}, the bound on the reserve after a jump, the dynamic program that chooses the reset instants, a remark on the certificate, and additional results of the case studies, with the notation and the equation numbers of the paper.

\section{Proof of Theorem~\ref{thm:cert}}
Suppose that the relay trips at some $t^\star\le T$. It then picked up at $s=t^\star-T_i^{\del}\ge0$, so $v_{ij}(s;\m q)\ge V_i^{\pk}$, and never reset, so $v_{ij}(t;\m q)\ge r_iV_i^{\pk}$ on $[s,t^\star]$. Let $t_n$ be the last reset instant with $t_n\le s$. If $n<N$ and $t_{n+1}\le t^\star$, the reset instant $t_{n+1}$ lies in $(s,t^\star]$, and~\eqref{eq:cert_reset} contradicts the hold. Otherwise $s\le t_{n+1}-T_i^{\del}$, and~\eqref{eq:cert_pickup} contradicts the pickup at $s$. For the converse, take as reset instants the times at which the relay with both levels lowered by more than $\delta_i$ resets after each of its pickups. This relay does not trip, so each of its pickups precedes the next reset instant by less than $T_i^{\del}$, which gives~\eqref{eq:cert_reset}, and its voltage is below its lowered pickup level outside these episodes, which gives~\eqref{eq:cert_pickup}. \hfill$\blacksquare$

\section{Proof of Proposition~\ref{prop:necessary}}
Every control has a positive latency, so $v_{ij}(0;\m q)=v_i^0+p_{ij}(0)\ge V_i^{\pk}$ and the relay picks up at $t=0$. It trips at $T_i^{\del}$ unless its voltage falls below the reset level at some $t\in(0,T_i^{\del}]$, that is, unless $\sum_kg_{ik}(t)q_k>p_{ij}(t)-h_i^{\rs}$ for some such $t$. Because $q_k\ge0$, the left side is at most $\sum_k\bar g_{ik}q_k$ and the right side is at least $\underline p_{ij}-h_i^{\rs}$, which gives~\eqref{eq:necessary}. Its left side is nondecreasing in $\m q$, so if it fails at $\m q^{\max}$, it fails for every $\m0\le\m q\le\m q^{\max}$. \hfill$\blacksquare$

\section{Proof of Proposition~\ref{prop:one}}
The relay trips in $[0,T]$ if and only if there is an $s\in[0,T-T_i^{\del}]$ at which it is above pickup, $p_{ij}(s)-g_{ik}(s)q_k\ge h_i$, and from which it stays above the reset level for the delay, $p_{ij}(t)-g_{ik}(t)q_k\ge h_i^{\rs}$ on $[s,s+T_i^{\del}]$. With the conventions of~\eqref{eq:a}, the first condition is $q_k\le a_{ij}^{\pk}(s)$ and the second is $q_k\le a_{ij}^{\rs}(t)$ on the interval, so the relay trips if and only if $q_k$ is at most the inner minimum of~\eqref{eq:qstar} for some $s$, which is the claim. A larger delay shrinks the range of $s$ and extends every interval of the inner minimum, so $q_{ij}^\star$ cannot increase. A larger latency shifts the unit response to the right, which cannot increase $g_{ik}(t)$ at any $t$ when the response is nondecreasing. For $q_k\ge0$, this cannot decrease $v_{ij}(t;q_k)$ at any $t$, so every action that lets the relay trip with the shorter latency lets it trip with the longer one, and the reserve cannot decrease. \hfill$\blacksquare$

\section{Bound on the Reserve After a Jump}
If event $j$ removes the reactive power absorption $Q_j^{\abs}$, $S_{ij}$ is the sensitivity of the transmission-side voltage of plant $i$ to this loss, and $n_i$ is the turns ratio between the transmission side and the relay side, then~\cite{albustami2026iberian}
\begin{equation}\label{eq:jump}
p_{ij}(0)=S_{ij}\,Q_j^{\abs}/n_i .
\end{equation}
\begin{corollary}[Bound from the jump]\label{cor:algebraic}
If $p_{ij}(0)\ge h_i$ and $p_{ij}(t)\ge p_{ij}(0)$ for $0<t\le T_i^{\del}$, then
\begin{equation}\label{eq:algebraic}
q_{ij}^\star\ \ge\ \big(S_{ij}\,Q_j^{\abs}-n_ih_i^{\rs}\big)/\bar S_{ik},
\end{equation}
where $\bar S_{ik}=n_i\bar g_{ik}$ is the largest reduction of the relay voltage of plant $i$ per MVAr of control $k$ within the delay, referred to the transmission side.
\end{corollary}
\begin{IEEEproof}
Take $s=0$ in~\eqref{eq:qstar}. Then $a_{ij}^{\pk}(0)=+\infty$, because $g_{ik}(0)=0$ and $p_{ij}(0)\ge h_i$. For $0<t\le T_i^{\del}$, the numerator of $a_{ij}^{\rs}(t)$ is at least $p_{ij}(0)-h_i^{\rs}\ge h_i-h_i^{\rs}\ge0$ and its denominator is at most $\bar g_{ik}$, so $a_{ij}^{\rs}(t)\ge(p_{ij}(0)-h_i^{\rs})/\bar g_{ik}$. Substituting~\eqref{eq:jump} and $\bar g_{ik}=\bar S_{ik}/n_i$ gives~\eqref{eq:algebraic}.
\end{IEEEproof}
The bound grows with the sensitivity $S_{ij}$ and the lost reactive power absorption, decreases as the margin to the reset level referred to the transmission side, $n_ih_i^{\rs}$, grows, and is inversely proportional to the effect of the control at the relay.

\section{Choice of the Reset Instants}
Algorithm~\ref{alg:reserve} chooses the reset instants of each pair along the current action of its event. Let $\m d$ be this action and consider the actions $s\m d$ with $s\ge0$. With a time step $\Delta$ and $t_n=n\Delta$, let $a_n^{\rs}$ be the multiple $s$ above which~\eqref{eq:cert_reset} holds at $t_n$ and $a_n^{\pk}$ the multiple above which~\eqref{eq:cert_pickup} holds on $[t_n,t_{n+1}]$, computed as in~\eqref{eq:a} with the bounds between time steps. With $m=T_i^{\del}/\Delta$, condition~\eqref{eq:cert_pickup} is needed on the interval $[t_n,t_{n+1}]$ when no reset instant lies in $(t_n,t_n+T_i^{\del}]$ and $t_{n+1}\le T-T_i^{\del}$. The dynamic program
\begin{equation}\label{eq:dp}
F_c=\max\Big\{a_c^{\rs},\min_{0\le b<c}\max\Big\{F_b,\!\max_{b\le n\le c-m-1}\!a_n^{\pk}\Big\}\Big\},\ F_0=-\infty,
\end{equation}
gives the smallest multiple $F_c$ that a set of reset instants whose last instant is $t_c$ needs. The reset instants of the $c$ that minimizes $\max\{F_c,\max_{c\le n\le n_T}a_n^{\pk}\}$ are kept, where $n_T$ is the last interval that ends by $T-T_i^{\del}$ and $c=0$ means no reset instant. For one control with $g_{ik}\ge0$, this program returns~\eqref{eq:qstar} at the time steps with both levels lowered by $\delta_i$. Algorithm~\ref{alg:reserve} alternates this choice with stages~\eqref{eq:stage1} and~\eqref{eq:stage2} until the result stops improving, where a result is better if it leaves fewer pairs uncovered, then if its total slack is smaller, and then if its reserve is smaller. It starts from the action of the pickup certificate and once more from reset instants spaced closer than the delay, and it keeps the better result.

\begin{algorithm}[t]
	\caption{Reserve LP with selection of the reset instants}\label{alg:reserve}
	\begin{algorithmic}[1]
		\Require responses $p_{ij}(t)$ and $g_{ik}(t)$, relay settings, allowances $\delta_i$, capabilities $\m q^{\max}$, and the sets $\mathcal R_j$, $\mathcal S_j$, $\mathcal K_j$
		\Ensure reserve $\bar{\m q}$, corrective actions $\m q^{(j)}$, and covered pairs
		\State $\m d^{(j)}\gets$ action of the pickup certificate, or $\m q^{\max}$ on $\mathcal K_j$ if this action is zero
		\Repeat
		\State choose the reset instants of every pair $(i,j)$ along $\m d^{(j)}$ by~\eqref{eq:dp}
		\State solve~\eqref{eq:stage1} and~\eqref{eq:stage2}, and set $\m d^{(j)}\gets\m q^{(j)}$
		\Until{the result stops improving, or after six passes}
		\State repeat lines 2--5 from reset instants spaced closer than $T_i^{\del}$ and keep the better result
		\State solve~\eqref{eq:stage3} for every event $j$, choosing its reset instants as in lines 2--5
		\State check every pair with $\eta_{ij}=0$ against~\eqref{eq:cert}
	\end{algorithmic}
\end{algorithm}

\section{Remark on the Certificate}
\begin{remark}\label{rem:cert}
	When consecutive reset instants are closer than $T_i^{\del}$, the certificate checks fewer than $T/T_i^{\del}+1$ instants, between which the relay may pick up. A relay with a large hysteresis is pulled below its reset level once after a pickup and only held below pickup afterwards. When $T_i^{\del}$ is uncertain, the certificate for the smallest possible delay holds for every larger one. When a latency is uncertain and the response of the control is nondecreasing after it, the certificate for the largest possible latency holds for every smaller one.
\end{remark}

\section{Additional Results of the Case Studies}
\textit{Systems and reserves.} Table~\ref{tab:systems} gives the size of each system of Section~\ref{sec:cases}, its fast capability, and its reserve for both reset ratios, both delays, and an error of 0.005~pu in the relay voltage.

\begin{table}[t]
\centering
\caption{Systems and reserves of Section~\ref{sec:cases} with $T=2$~s and $\delta_i=0.05h_i$. The fast capability excludes manually switched reactors. Reserves are in MVAr and cover every pair at risk except where marked. The last three rows lower every margin by 0.005~pu, an error of the relay voltage, with reset at pickup. $^\ast$One pair is uncovered.}
\label{tab:systems}
\footnotesize
\setlength{\tabcolsep}{4pt}
\begin{tabular}{@{}lrrrr@{}}
\toprule
 & IEEE 39 & NPCC & GB & 2000 bus \\
\midrule
Plants & 12 & 16 & 20 & 26 \\
Contingencies in the list & 45 & 54 & 68 & 74 \\
Events with a pair at risk & 12 & 25 & 25 & 46 \\
Pairs at risk & 28 & 53 & 69 & 399 \\
Fast capability (MVAr) & 753 & 2034 & 3176 & 6129 \\
\multicolumn{5}{@{}l}{\textit{Reserve}} \\
Reset at pickup, 0.16~s & 48 & 491 & 610 & 608 \\
Reset at pickup, 1~s & 37 & 180 & 490 & 527 \\
Reset 0.98, 0.16~s & 187 & 677 & 924 & 627 \\
Reset 0.98, 1~s & 138 & 525 & 758 & 586 \\
\multicolumn{5}{@{}l}{\textit{With a relay voltage error of 0.005~pu}} \\
Pairs at risk & 58 & 71 & 135 & 539 \\
Reserve, 0.16~s & 70 & 586 & 1875$^\ast$ & 633 \\
Reserve, 1~s & 55 & 282 & 1660 & 559 \\
\bottomrule
\end{tabular}
\end{table}

\begin{table*}[t]
\centering
\caption{Response-based plant rules against the LP actions with $T=15$~s, simulated for 30~s with 0.16-s and 1-s relays that reset at pickup, for every event with a pair at risk (12 on IEEE 39, 25 on NPCC). Trips are summed over the events, and those after $T$ follow a plus sign. MVAr is the reactive power in use 2~s after the event, summed over the events, and $V_{\min}$ is the lowest bus voltage in pu. No run collapses. On NPCC no event is certified with $T=15$~s, so the LP actions and reserve are zero.}
\label{tab:laws}
\footnotesize
\setlength{\tabcolsep}{3.2pt}
\begin{tabular}{@{}lrrrrrrrrrrrr@{}}
\toprule
 & \multicolumn{6}{c}{IEEE 39} & \multicolumn{6}{c}{NPCC} \\
\cmidrule(lr){2-7}\cmidrule(l){8-13}
 & \multicolumn{3}{c}{0.16~s} & \multicolumn{3}{c}{1~s} & \multicolumn{3}{c}{0.16~s} & \multicolumn{3}{c}{1~s} \\
\cmidrule(lr){2-4}\cmidrule(lr){5-7}\cmidrule(lr){8-10}\cmidrule(l){11-13}
Control & Trips & MVAr & $V_{\min}$ & Trips & MVAr & $V_{\min}$ & Trips & MVAr & $V_{\min}$ & Trips & MVAr & $V_{\min}$ \\
\midrule
LP action & 39 & 264 & 0.95 & 23 & 226 & 0.96 & 90 & 0 & 0.82 & 26+3 & 0 & 0.91 \\
Full absorption, 100~ms after own pickup & 0 & 411 & 1.00 & 0 & 411 & 1.00 & 0 & 2541 & 0.89 & 0 & 2541 & 0.89 \\
Same, limited to the LP reserve & 30 & 246 & 0.95 & 22 & 204 & 0.96 & 90 & 0 & 0.82 & 26+3 & 0 & 0.91 \\
Full absorption, 1~s after own pickup & 71 & 1694 & 0.95 & 31 & 658 & 0.96 & 90 & 7432 & 0.82 & 11 & 3229 & 0.90 \\
Full absorption, 100~ms after high side above 1.10~pu & 64 & 1029 & 0.95 & 28 & 273 & 0.97 & 90 & 1294 & 0.82 & 26+3 & 0 & 0.91 \\
Voltage droop & 7 & 503 & 1.00 & 7 & 463 & 1.00 & 9 & 2257 & 0.91 & 2 & 1807 & 0.94 \\
\bottomrule
\end{tabular}
\end{table*}

\textit{Time grid, runtime, and exact programs.} The classes of Fig.~\ref{fig:delay}(a) are computed with a time step of 5~ms. Three quarters of the labels that require preventive action or are not correctable are proven in continuous time, and about half of the others change with a time step of 1~ms, most of them to correctable. The LP of the 2000-bus system solves in at most 128~s on one core. On IEEE 39, the action of each event equals the exact optimum of a mixed-integer program with the relay logic in all 28 event problems solved to optimality, and the exact program for the reserve of all events found no better value within one hour without proving optimality.

\textit{Reactor rounding.} Rounding every reactor action up to its rating, as a mechanically switched bank would need, raises the reserve only on the 2000-bus system, by 82~MVAr with 0.16-s relays, and keeps every pair certified.

\textit{Latency and location.} Fig.~\ref{fig:latency} gives the curves behind the numbers of Section~\ref{sec:cases_insights}.

\begin{figure}[t]
\centering
\includegraphics[width=\columnwidth]{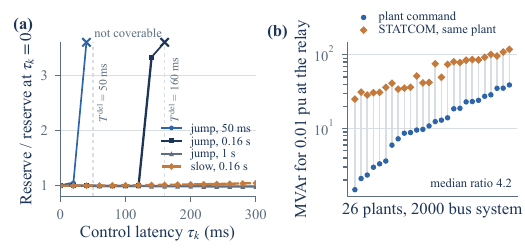}
\caption{Effect of the latency and of the location. (a) Reserve of plant set points versus the control latency, relative to zero latency, for the load disconnection of Fig.~\ref{fig:validation}(a) (jump) and the Seville plant trip of the 2000-bus system (slow). (b) MVAr that lowers the relay voltage of each plant by 0.01~pu, from its own set point and from a STATCOM at its transmission bus.}
\label{fig:latency}
\end{figure}

\textit{Tolerance of the smallest actions.} The smallest actions leave little room. They reset the relays 12 to 152~ms before the delay ends, a tenth to a half less magnitude brings a trip back in all 14 cases tested, and 20 to 80~ms more latency does so in 11 of them.

\textit{Actions held for 30~s.} Of the 21 actions simulated for 30~s, five let plants trip after the window of 2~s. In four load disconnections of IEEE 39 the voltage keeps rising and the first plants trip between 2.12 and 2.54~s, and after one generator trip on the 2000-bus system plants trip later and the simulation stops near 21~s. Fig.~\ref{fig:window}(a) shows one of them, the disconnection of load PQ\_6 of IEEE 39 with 0.16-s relays. Under the action of 12.7~MVAr sized with $T=2$~s, the relay of plant RES12 picks up briefly and resets, but its voltage keeps rising, and the relay picks up again at 2.38~s and trips at 2.54~s. Nine more plants trip by 3.1~s. The action of 21.7~MVAr sized with $T=15$~s and a reset instant at 15~s keeps the voltage below pickup after the first 0.12~s, and no plant trips in 30~s.

\begin{figure}[t]
\centering
\includegraphics[width=\columnwidth]{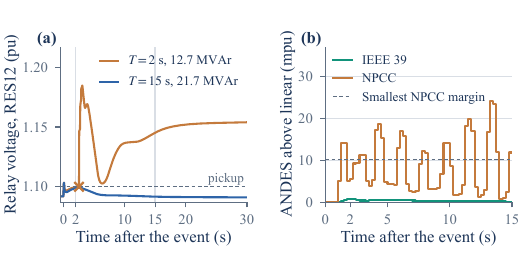}
\caption{Actions held for 30~s and error of the linear model. (a) Relay voltage of plant RES12 after the disconnection of load PQ\_6 of IEEE 39 with 0.16-s relays, under the actions sized with $T=2$~s and $T=15$~s and held for 30~s. The cross marks the trip, and the vertical lines mark 2 and 15~s. (b) Largest amount by which the relay voltage simulated without action and with the relays disabled exceeds the linear model, over every event and relay, in intervals of 0.25~s.}
\label{fig:window}
\end{figure}

Fig.~\ref{fig:window}(b) shows how far the simulation without action rises above the linear model over this longer window. On IEEE 39 the excess stays below 0.001~pu. On NPCC it exceeds the smallest margin of 0.010~pu in several intervals after 1~s and reaches 0.024~pu. Six NPCC relays then have a margin below the allowance and cannot be certified, and every action that covers the remaining pairs raises one of these relays, so no NPCC event is certified with $T=15$~s.

\textit{Response-based rules.} Table~\ref{tab:laws} compares the rule of Section~\ref{sec:cases_scheme}, in which each plant applies its full absorption 100~ms after its own relay picks up, with the LP actions and three other response-based rules on IEEE 39 and NPCC. On IEEE 39 the LP actions leave trips only in the five events that hold a pair the certificate cannot cover, at both delays. The other rules leave trips at both delays. They are the full absorption applied 1~s after the plant's own relay picks up, the full absorption applied 100~ms after the high-side voltage of the main power transformer exceeds 1.10~pu, and a voltage droop on the collector voltage. The rule of Section~\ref{sec:cases_scheme} also leaves trips in 5 of the 12 events of IEEE 39 when each plant is limited to the reserve that the LP holds for it.
\fi

\end{document}